\documentstyle[12pt]{ioplppt}

\begin{document}

\title{Polynomial rings of the chiral $SU(N)_2$ models}
\author{A Lima-Santos}

\address{Departamento de F\' {i}sica, Universidade Federal de
 S\~ao Carlos, Caixa Postal 676, 13569-905 S\~ao Carlos, Brazil}

\begin{abstract}
Via explicit diagonalization of the chiral $SU(N)_{2}$ fusion matrices, we
discuss the possibility of representing the fusion ring of the chiral $SU(N)$
models, at level $K=2$, by a polynomial ring in a single variable when $N$
is odd and by a polynomial ring in two-variable when $N$ is even.
\end{abstract}


\section{Introduction}

Six years ago , Gepner conjectured that the fusion ring of theories with $%
SU(N)$ current algebra is isomorphic to a ring in $N-1$ variables associated
to the fundamental representations, quotiented by an ideal of constraints
that derive with a potential \cite{Gepner2}.

Four years ago, Di Francesco and Zuber postulated a necessary and sufficient
condition for a one-variable polynomial ring \cite{Zuber}: Assume that among
the matrices $N_{i},i=1,...,n$ , there exists at least one, call it $N_{f}$
, with non degenerated eigenvalues. Thus, any other $N_{i}$ may be
diagonalized in the same basis as $N_{f}$ and there exists a unique
polynomial $P_{i}(x)$ of degree at most $n-1$ such that its eigenvalues $%
\gamma _{i}^{(l)}$ satisfy 
\begin{equation}
\gamma _{i}^{(l)}=P_{i}(\gamma _{f}^{(l)})  \label{eq0.1}
\end{equation}
$P_{i}$ being given by the Lagrange interpolation formula. Therefore, any $%
N_{i}$ may be written as 
\begin{equation}
N_{i}=P_{i}(N_{f})  \label{eq0.2}
\end{equation}
with a polynomial $P_{i}$ ; as both $N_{i}$ and $N_{f}$ have integral
entries, $P_{i}(x)$ must have rational coefficients.

The $n\times n$ matrix $N_{f}$ , on the other hand, satisfies its
characteristic equation ${\cal P}(x)=0$ , that is also its minimal equation,
as $N_{f}$ has no degenerate eigenvalues. The constraint on $N_{f}$ is thus 
\begin{equation}
{\cal P}(N_{f})=0  \label{eq0.3}
\end{equation}
that may of course be integrated to yield a ''potential '' ${\cal W}(x)$ ,
which is a polynomial of degree $n+1$. In this way, Di Francesco and Zuber
have characterized the rational conformal field theories (RCFTs) which have
a description in terms of a fusion potential in one variable. Moreover, they
have also proposed a generalized potential to describe other theories. In
reference\cite{Ofer} Aharony have determined a simple criterion to a
generalized description of RCFTs by fusion potentials in more than one
variable.

In this note we tackle this problem and discuss the possibility of
representing the fusion rings of the chiral $SU(N)_{2}$ models, by
polynomial rings in two variables. Exploiting the Di Francesco and Zuber
condition we show that these polynomial rings in two variables are reduced
to polynomial rings in a single variable in the cases for which $N$ is odd
(or $N=2$ ).

In section 2 we discuss some algebraic setting of the chiral RCFTs. Section
3 describes the primary fields of the chiral $SU(N)$ models, at level $K=2$,
in {\em cominimal equivalence classes}. In the last section we report a
computer study which diagonalizes the fusion matrices of the chiral $%
SU(N)_{2}$ models and gives their polynomial rings in one and two variables.

\section{Fusion algebras}

Fusion algebras are found to play an important role in the study of RCFTs.
Beside the fact that the fusion rules can be expressed in terms of the
unitary matrix $S$ \cite{Verlinde} that encodes the modular transformations
of the characters of the RCFT 
\begin{equation}
N_{ij}^{k}=\sum_{l}\frac{S_{il}}{S_{0l}}\ S_{jl}\ S_{kl}^{*}.  \label{eq1.1}
\end{equation}
Here ''$0$ '' refers to the identity operator, and the labels $i,...,l$ run
over $n$ values corresponding to the primary fields of the chiral algebra of
the RCFT. There is a more fundamental reason to look for representations of
the fusion algebra, based on the concept of operator products\cite{BPZ}.
When one tries to compute the operator product coefficients, one is almost
inevitably led to the concept of fusion rules, i.e. formal products 
\begin{equation}
A_{i}\ A_{j}=\sum_{k}N_{ij}^{k}\ A_{k}.  \label{eq1.2}
\end{equation}
of primary fields describing the basis-independent content of the operator
product algebra.

By definition, the fusion rule coefficients possess the property of
integrality $N_{ij}^{k}\in Z_{\geq 0}$. In addition, they inherit several
simple properties:

\begin{itemize}
\item  {\it symmetry}: $N_{ij}^{k}=N_{ji}^{k}$;

\item  {\it associativity}: $\sum_{k}N_{ij}^{k}N_{kl}^{m}=%
\sum_{k}N_{jl}^{k}N_{ik}^{m}$;

\item  {\it existence of unit}: there is an index ''$0$ '' (identity
operator) such that $N_{i0}^{j}=\delta _{i}^{j}$ and

\item  {\it charge conjugation}: $N_{ijl}=%
\sum_{k}N_{ij}^{k}C_{kl}=(N_{ij}^{l})^{\dagger }$ is completely symmetric in
the indices $i,j,l$.
\end{itemize}

Because of these properties, one can interpret the fusion rule coefficients
as the structure constants of a commutative associative ring with basis
given by the primary fields.

The matrix $S$ implements the modular transformation $\tau \rightarrow
-1/\tau $ and obeys $S^{2}=C$. In addition, the diagonal matrix $T_{ii}=\exp
(2i\pi (\Delta _{i}-c/24))$, where $\Delta _{i}$ is the conformal dimension
of the primary field $i$ and $c$ is the central charge, implements the
modular transformation $\tau \rightarrow \tau +1$ and obeys $(ST)^{3}=C$,
which implies a relation between the structure constants $N_{ij}^{k}$ and
the conformal dimensions $\Delta _{i}$ \cite{Vafa}: 
\begin{equation}
N_{ijkl}(\Delta _{i}+\Delta _{j}+\Delta _{k}+\Delta
_{l})=\sum_{r}N_{ijklr}\Delta _{r}  \label{eq1.2a}
\end{equation}
where 
\begin{equation}
N_{ijkl}=N_{ij}^{\stackrel{\_}{n}}N_{kl}^{n}\quad \text{and\quad }%
N_{ijklr}=N_{ij}^{r}N_{klr}+N_{jk}^{r}N_{ilr}+N_{ik}^{r}N_{jlr}
\label{eq1.2b}
\end{equation}

It was suggested in \cite{GK} that these proprieties fully characterize a
RCFT , and that any commutative ring satisfying these properties is the
fusion ring of some RCFT.

The matrices $N_{i}$ defined by $(N_{i})_{jk}=N_{ij}^{k}$ form themselves a
trivial representation of the fusion algebra 
\begin{equation}
N_{i}\ N_{j}=\sum_{k}N_{ij}^{k}\ N_{k}  \label{eq1.3}
\end{equation}
as follows from unitarity of the matrix $S$; this expresses the
associativity property of the algebra (\ref{eq1.2}). The relation (\ref
{eq1.1}) implies that the matrix $S$ diagonalizes the matrices $N_{i}$ and
their eigenvalues are of the form 
\begin{equation}
\gamma _{i}^{(l)}=\frac{S_{il}}{S_{0l}}  \label{eq1.4}
\end{equation}
and obey the sum rules 
\begin{equation}
\gamma _{i}^{(l)}\ \gamma _{j}^{(l)}=\sum_{k}N_{ij}^{k}\ \gamma _{k}^{(l)}
\label{eq1.5}
\end{equation}

The general study of these fusion algebras and their classification have
been the object of much work \cite{GK}-\cite{CPR2}.

The numbers 
\begin{equation}
d_{i}\doteq \gamma _{i}^{(0)}=\frac{S_{i0}}{S_{00}}  \label{eq1.6}
\end{equation}
appear as statistical dimensions of superselection sectors \cite{R1},\cite
{R3} in algebraic quantum field theory; as square roots of indices for
inclusions of von Neumann algebras \cite{R2}; as relative sizes of highest
weight modules of chiral symmetry algebras in conformal field theory \cite
{Verlinde}; and in connection with truncated tensor products of quantum
groups (see \cite{Fuchs} for an accomplished review). According to (\ref
{eq1.5}), these numbers obey the statistical dimension sum rules 
\begin{equation}
d_{i}\ d_{j}=\sum_{k}N_{ij}^{k}\ d_{k}.  \label{eq1.7}
\end{equation}
which shows that $d_{i}$ is a Frobenius eigenvalue of $N_{i}$.

\section{$SU(N)_{2}$ cominimal equivalence classes}

At the level $K=2$ the central charge of the chiral $SU(N)$ models is given
by 
\begin{equation}
c=\frac{2(N-1)}{N+2}  \label{eq1.17a}
\end{equation}
and their primary fields are identified with the order fields $\sigma _{k}$
, $k=0,1,...,N-1$; $Z_{N}$-neutral fields $\epsilon ^{(j)},j=1,2,...\leq N/2$
and the parafermionic currents $\Psi _{k}$, $k=1,...,N-1$ , in
Zamolodchikov-Fateev's parafermionic theories\cite{Zamo1}. For each primary
field we define a ''charge '' $\nu =0,1,...,2(N-1)\quad \bmod\ 2N$ and we
collect the $N(N+1)/2$ primary fields in $N$ {\em cominimal equivalence
classes }\cite{NRS}, $[\phi _{k}^{k}]$ , $k=0,1,...,N-1,$ according to their
statistical dimensions: 
\begin{eqnarray}
d_{k} &=&\prod_{i=0}^{k-1}\frac{s(N-i)}{s(i+1)},\qquad s(x)=\sin (\frac{x\pi 
}{N+2})  \nonumber \\
d_{0} &=&1,\qquad d_{N-k}=d_{k},\qquad k=1,2,...,N-1  \label{eq1.17}
\end{eqnarray}

$SU(N)_{2}$ representations of the order fields $\phi _{k}^{k}$ , $%
k=1,...,N-1$ are the fully antisymmetric Young tableaux with $k$ boxes (i.e.
the reduced tableau which is a column with $k$ boxes). Tableaux of fields
comprising a cominimal equivalence class $\phi _{\nu }^{k}$ in which the
representation $\phi _{k}^{k}$ appears, $(\nu =k\ \bmod\ {2}$ , i.e., $\nu
=k,k+2,\cdots ,2N-2-k)$, are obtained by adding $(\nu -k)/2$ rows of width $%
2 $ to the top of the reduced tableau of $\phi _{k}^{k}$ . Therefore $\phi
_{\nu }^{k}$ is a Young tableau of two columns with $\nu $ boxes, since $%
(\nu +k)/2$ boxes in the first column and $(\nu -k)/2$ in the second column.

The conformal weights of the fields comprising a cominimal equivalence class
in which the representation $\phi _{k}^{k}$ appears are simply related to
the conformal weight of $\phi _{k}^{k}$ by 
\begin{equation}
\Delta _{\nu }^{k}=\Delta _{k}^{k}+\frac{\nu -k}{4N}(2N-\nu -k)
\label{eq1.17b}
\end{equation}
and the conformal dimensions of the order fields \cite{Zamo1} are given by 
\begin{equation}
\Delta _{k}^{k}=\frac{k(N-k)}{2N(N+2)}  \label{eq1.17c}
\end{equation}

These equivalence classes are generated by $Z_{N}$ symmetry which connect
the representations belonging to each class through of the fusion rules\cite
{Gepner1} 
\begin{equation}
\phi _{\nu _{1}}^{k_{1}}\times \phi _{\nu
_{2}}^{k_{2}}=\sum_{k=|k_{1}-k_{2}|\bmod\ {2}}^{\min
(k_{1}+k_{2},2N-k_{1}-k_{2})}\phi _{\nu _{1}+\nu _{2}}^{k}  \label{eq1.18}
\end{equation}
In particular, the elementary field $\phi _{1}^{1}$, $(\phi _{1}^{1}\times
\phi _{\nu }^{k}=\phi _{\nu +1}^{k-1}+\phi _{\nu +1}^{k+1})$ connects the
equivalence class of $\phi _{\nu }^{k}$ with adjacent classes, while the
field $\phi _{2}^{0}$ , $(\phi _{2}^{0}\times \phi _{\nu }^{k}=\phi _{\nu
+2}^{k})$, connects the fields in the same cominimal equivalence class.
Thus, the $SU(N)_{2}$ fusion ring can be generated by these two fields. For
example, the $10$ primary fields of $SU(4)_{2}$ can be collected in $4$
cominimal equivalence classes as 
\begin{equation}
\left\{ 
\begin{array}{llllllllll}
&  &  & \phi _{3}^{3} &  &  &  & _{\rightarrow } & d_{3}=\frac{s(2)}{s(1)} & 
\\ 
&  & \quad \nearrow & \quad \searrow &  &  &  &  &  &  \\ 
&  & \phi _{2}^{2} &  & \phi _{4}^{2} &  &  & \rightarrow & d_{2}=\frac{s(3)%
}{s(1)} &  \\ 
& \quad \nearrow & \quad \searrow & \quad \nearrow & \quad \searrow &  &  & 
&  &  \\ 
& \phi _{1}^{1} &  & \phi _{3}^{1} &  & \phi _{5}^{1} &  & \rightarrow & 
d_{1}=\frac{s(4)}{s(1)} &  \\ 
\quad \nearrow & \quad \searrow & \quad \nearrow & \quad \searrow & \quad
\nearrow & \quad \searrow &  &  &  &  \\ 
\phi _{0}^{0} &  & \phi _{2}^{0} &  & \phi _{4}^{0} &  & \phi _{6}^{0} & 
\rightarrow & d_{0}=\frac{s(5)}{s(1)} & 
\end{array}
\right\}  \label{eq1.19}
\end{equation}
These cominimal equivalence classes provide a representation of the $Z_{4}$
symmetry and the primary fields corresponding to representations in the same
class differ only by free fields.

\section{$SU(N)_{2}$ polynomial rings}

Let us start by considering the case $SU(4)_{2}$ (the $SU(2)_{2}$ and $%
SU(3)_{2}$ cases were considered in \cite{Zuber}):

The variables $x$ and $y$ are associated to the fields $\phi _{1}^{1}$ and $%
\phi _{2}^{0}$ , respectively.. Using $\phi _{0}^{0}=1$ , the fusion rules (%
\ref{eq1.18}) (see table(\ref{eq1.19})) give the expressions of the other
fields

\medskip

\begin{equation}
\begin{array}{lllllll}
\phi _{0}^{0}=1 &  & \phi _{1}^{1}=x &  & \phi _{2}^{2}=x^{2}-y &  & \phi
_{3}^{3}=x^{3}y-2xy \\ 
\phi _{2}^{0}=y &  & \phi _{3}^{1}=xy &  & \phi _{4}^{2}=x^{2}y-y^{2} &  & 
\\ 
\phi _{4}^{0}=y^{2} &  & \phi _{5}^{1}=xy^{2} &  &  &  &  \\ 
\phi _{6}^{0}=y^{3} &  &  &  &  &  & 
\end{array}
\label{eq1.20}
\end{equation}
and from the identification $\phi _{\nu }^{k}=\phi _{4+\nu }^{4-k}$ $\bmod\
8 $ we get the following constraints 
\begin{equation}
\begin{array}{lll}
x^{4}-3x^{2}y+y^{2}=1 &  & x^{3}y-2xy^{2}=x \\ 
x^{2}y^{2}-y^{3}=x^{2}-y &  & x^{3}-2xy=xy^{3} \\ 
y^{4}=1 &  & 
\end{array}
\label{eq1.21}
\end{equation}
These constraints can be combined and reduced to a one-variable constraint 
\begin{equation}
x^{10}-8x^{6}-9x^{2}=0,  \label{eq1.22}
\end{equation}
which is equal to the characteristic equation of the fusion matrix $N_{\phi
_{1}^{1}}$ , and its eigenvalue $0$ is doubly degenerate implying that $x$
may not be inverted on the ring. Similarly, one can eliminate $x$ from (\ref
{eq1.21}) and get a one-variable constraint in $y$%
\begin{equation}
y^{10}-y^{8}-2y^{6}+2y^{4}+y^{2}-1=0  \label{eq1.23}
\end{equation}
which is equal to the characteristic equation of the fusion matrix $N_{\phi
_{2}^{0}}$ , whose eigenvalues are degenerate. Thus, the fusion ring of the $%
SU(4)_{2}$ model can be expressed in terms of two variables associated with
the representations $\phi _{1}^{1}$ and $\phi _{2}^{0}$ which satisfy
independent constraint equations.

Next, let us consider the $15$ primary fields of the chiral $SU(5)_{2}$
model which can be collected in $5$ cominimal equivalence classes as: 
\begin{equation}
\fl \left\{ 
\begin{array}{lllllllllll}
&  &  &  & \phi _{4}^{4} &  &  &  &  & \rightarrow & d_{4}=\frac{s(2)}{s(1)}
\\ 
&  &  & \quad \nearrow & \quad \searrow &  &  &  &  &  &  \\ 
&  &  & \phi _{3}^{3} &  & \phi _{5}^{3} &  &  &  & \rightarrow & d_{3}=%
\frac{s(3)}{s(1)} \\ 
&  & \quad \nearrow & \quad \searrow & \quad \nearrow & \quad \searrow &  & 
&  &  &  \\ 
&  & \phi _{2}^{2} &  & \phi _{4}^{2} &  & \phi _{6}^{2} &  &  & \rightarrow
& d_{2}=\frac{s(4)}{s(1)} \\ 
& \quad \nearrow & \quad \searrow & \quad \nearrow & \quad \searrow & \quad
\nearrow & \quad \searrow &  &  &  &  \\ 
& \phi _{1}^{1} &  & \phi _{3}^{1} &  & \phi _{5}^{1} &  & \phi _{7}^{1} & 
& \rightarrow & d_{1}=\frac{s(5)}{s(1)} \\ 
\quad \nearrow & \quad \searrow & \quad \nearrow & \quad \searrow & \quad
\nearrow & \quad \searrow & \quad \nearrow & \quad \searrow &  &  &  \\ 
\phi _{0}^{0} &  & \phi _{2}^{0} &  & \phi _{4}^{0} &  & \phi _{6}^{0} &  & 
\phi _{8}^{0} & \rightarrow & d_{0}=\frac{s(6)}{s(1)}
\end{array}
\right\}  \label{eq1.24a}
\end{equation}
The variables $x$ and $y$ are associated to the fields $\phi _{1}^{1}$ and $%
\phi _{2}^{0}$ , respectively. Using $\phi _{0}^{0}=1$ , the fusion rules (%
\ref{eq1.18}) give the expressions of the other fields 
\begin{equation}
\begin{array}{lllll}
\phi _{0}^{0}=1 &  & \phi _{1}^{1}=x &  & \phi _{4}^{2}=x^{2}y-y^{2} \\ 
\phi _{2}^{0}=y &  & \phi _{3}^{1}=xy &  & \phi _{6}^{2}=x^{2}y^{2}-y^{3} \\ 
\phi _{4}^{0}=y^{2} &  & \phi _{5}^{1}=xy^{2} &  & \phi _{3}^{3}=x^{3}-2xy
\\ 
\phi _{6}^{0}=y^{3} &  & \phi _{7}^{1}=xy^{3} &  & \phi
_{5}^{3}=x^{3}y-2xy^{2} \\ 
\phi _{8}^{0}=y^{4} &  & \phi _{2}^{2}=x^{2}-y &  & \phi
_{4}^{4}=x^{4}-3x^{2}y+y^{2}
\end{array}
\label{eq1.24b}
\end{equation}
and the identification $\phi _{\nu }^{k}=\phi _{5+\nu }^{5-k}$ $\bmod\ 10$
gives us the following constraint equations 
\begin{equation}
\begin{array}{lll}
x^{5}-4x^{3}y+3xy^{2}=1 &  & x^{2}y^{3}-y^{4}=x^{3}-2xy \\ 
x^{4}y-3x^{2}y^{2}+y^{3}=x &  & x^{4}-3x^{2}y+y^{2}=xy^{4} \\ 
x^{3}y^{2}-2xy^{3}=x^{2}-y &  & y^{5}=1
\end{array}
\label{eq1.x}
\end{equation}
These constraints can be combined and reduced to a one-variable constraint
equation 
\begin{equation}
x^{15}-16x^{10}-57x^{5}+1=0  \label{eq1.24c}
\end{equation}
which is equal to the characteristic equation of the fusion matrix $N_{\phi
_{1}^{1}}$ , whose eigenvalues are non-degenerate. It means that $x$ may be
inverted on the ring: we can eliminate $y$ from the constraint equations (%
\ref{eq1.x}) as: 
\begin{equation}
y=\frac{1}{181}(-14x^{12}+221x^{7}+910x^{2}).  \label{eq1.24d}
\end{equation}
Substituting this value of $y$ into (\ref{eq1.24b}) we will get a polynomial
ring in a single variable:

\[
\fl
\begin{array}{ll}
P_{0}^{0}(x)=1 & P_{1}^{1}(x)=x \\ 
&  \\ 
P_{2}^{0}(x)=\frac{1}{181}(910x^{2}+221x^{7}-14x^{12}) & P_{3}^{1}(x)=\frac{1%
}{181}(910x^{3}+221x^{8}-14x^{13}) \\ 
&  \\ 
P_{4}^{0}(x)=\frac{1}{181}(4592x^{4}\allowbreak +1260x^{9}-79x^{14}) & 
P_{5}^{1}(x)=\frac{1}{181}(79+89\allowbreak x^{5}-4x^{10}) \\ 
&  \\ 
P_{6}^{0}(x)=\frac{1}{181}(404x+155x^{6}-9x^{11}) & P_{7}^{1}(x)=\frac{1}{181%
}(404x^{2}+155x^{7}-9x^{12}) \\ 
&  \\ 
P_{8}^{0}(x)=\frac{1}{181}(2043x^{3}+597x^{8}-37x^{13}) & P_{3}^{3}(x)=-%
\frac{1}{181}(1639x^{3}+442x^{8}-28x^{13}) \\ 
&  \\ 
P_{2}^{2}(x)=-\frac{1}{181}(729x^{2}+221x^{7}-14x^{12}) & P_{5}^{3}(x)=-%
\frac{1}{181}(144+66x^{5}-5x^{10}) \\ 
&  \\ 
P_{4}^{2}(x)=-\frac{1}{181}(3682x^{4}+1039x^{9}-65x^{14}) & P_{4}^{4}(x)=%
\frac{1}{181}(2043x^{4}+597x^{9}-37x^{14}) \\ 
&  \\ 
P_{6}^{2}(x)=-\frac{1}{181}(325x+66x^{6}-5x^{11}) & 
\end{array}
\]
\vspace{0.25cm}{}

These $P_{\nu }^{k}(x)$ polynomials define ( modulo $%
x^{15}-16x^{10}-57x^{5}+1$) one-variable $SU(5)_{2}$ polynomial ring.

Similarly, one can eliminate $x$ from (\ref{eq1.x}) and get a one-variable
constraint in $y$%
\begin{equation}
y^{15}-3y^{10}+3y^{5}-1=0
\end{equation}
which is equal to the characteristic equation of the fusion matrix $N_{\phi
_{2}^{0}}$ , but their eigenvalues are degenerate.

We now extend this construction to the whole set of $SU(N)_{2}$ models. For
each irreducible representation $\phi _{\nu }^{k}$ we associate the
following polynomials

\begin{equation}
P_{\nu }^{k}(x,y)=\sum_{n=0}^{[\frac{k}{2}]}(-1)^{n}\frac{(k-n)!}{n!(k-2n)!}%
x^{k-2n}y^{n+\frac{\upsilon -k}{2}}  \label{eq1.24}
\end{equation}
where $k=0,1,...,N-1,\quad \nu =k\bmod\ {2,\ }{\rm i.e.\ }\nu
=k,k+2,...,2(N-1)-k$ and $[\frac{k}{2}]$ means the largest integer less than
or equal to $k/2$.

The identification $\phi _{\nu }^{k}=\phi _{N+\nu }^{N-k}$ $\bmod\ 2N$ gives
the corresponding one-variable constraint equations:

\begin{equation}
x^{\frac{N}{2}}\prod_{n=1}^{\frac{N}{2}}(x^{N}+(-1)^{n}d^{N}(n))=0,\qquad
(y^{\frac{N}{2}}-1)^{\frac{N+2}{2}}(y^{\frac{N}{2}}+1)^{\frac{N}{2}}=0\qquad
\label{eq1.25}
\end{equation}
for the cases when $N$ is even, and 
\begin{equation}
\prod_{n=1}^{\frac{N+1}{2}}(x^{N}-d^{N}(n))=0,\qquad (y^{N}-1)^{\frac{N+1}{2}%
}=0  \label{eq1.26}
\end{equation}
for the cases when $N$ is odd. In these expressions have introduced the
numbers

\begin{equation}
d(n) =\frac{\sin (n\pi \frac{N}{N+2})}{\sin (\frac{n\pi }{N+2})}, \quad n =
1,2,...,\leq \frac{N+2}2 .  \label{eq1.27}
\end{equation}

Inspecting the constraint equations in the variable $y$ we can see that the
fusion matrices $N_{\phi _{2}^{0}}$ are degenerate for all $SU(N)_{2}$
models. It means that we can not eliminate the variable $x$ from the
polynomials (\ref{eq1.24}). If $N$ is even and $N>2$, we can see from (\ref
{eq1.25}) that among the eigenvalues of fusion matrices $N_{\phi _{1}^{1}}$
only zero is degenerate ($N/2$ times), following that $x$ also can not be
inverted on these rings. It means that we also can not eliminate the
variable $y$ from (\ref{eq1.24}) and the corresponding fusion ring is
represented by a polynomial ring in two variables.

On the other hand, if $N$\ is odd or $N=2$, the eigenvalues of fusion
matrices $N_{\phi _{1}^{1}}$ are not degenerate and $x$ may be inverted on
the ring. We can therefore solve for $y$ as function of $x$ using the
corresponding constraint equations which were reduced to (\ref{eq1.26}) and
the fusion ring is faithfully represented by one variable polynomials. For
instance, the next $N$-odd models is $SU(7)_{2}$ for which the constraint
equation is $x^{28}-64x^{21}-157x^{14}+1640x^{7}+1=0$ and it is possible
eliminate $y$ from (\ref{eq1.24}) using: 
\begin{equation}
y=\frac{1}{664276}(2958x^{23}-189549x^{16}-4653716x^{9}+5504583x^{2}).
\label{eq1.31}
\end{equation}
and we get the resulting fusion ring as a polynomial ring in one variable.

At this point we can proceed to the generalization of these results by
explicit diagonalization of fusion matrices of the chiral $SU(N)_{2}$
models. To each irreducible representation $\phi _{\nu }^{k}$ we associate a
factored characteristic equation $\det (x{\bf 1}-N_{\phi _{\nu }^{k}})=0$
which depend on the parafermionic charge $\nu $ according to $N=\frac{p}{q}$ 
$\nu $ , where $p$ and $q$ are positive integers mutually coprime: 
\begin{equation}
\prod_{n=1}^{\frac{N+1}{2}}\left( x^{p}-d_{k}^{p}(n)\right) ^{\frac{\nu }{q}%
}=0,\text{ \quad {\rm if\quad }}p.q\text{{\rm -odd}}  \label{eq1.32}
\end{equation}
\begin{equation}
\prod_{n=1}^{\frac{N+1}{2}}\left( x^{p}+(-1)^{n}d_{k}^{p}(n)\right) ^{\frac{%
\nu }{q}}=0,\text{ \quad {\rm if\quad }}p.q\text{{\rm -even}}  \label{eq1.33}
\end{equation}
for $N$-odd, and 
\begin{equation}
\left( x^{p}-d_{k}^{p}(l)\right) ^{\frac{\nu }{2q}}\prod_{n=1}^{\frac{N}{2}%
}\left( x^{p}-d_{k}^{p}(n)\right) ^{\frac{\nu }{q}}=0,\text{ \quad {\rm %
if\quad }}p.q\text{{\rm -odd}}  \label{eq1.34}
\end{equation}
\begin{equation}
\left( x^{p}+(-1)^{l}d_{k}^{p}(l)\right) ^{\frac{\nu }{2q}}\prod_{n=1}^{%
\frac{N}{2}}\left( x^{p}+(-1)^{n}d_{k}^{p}(n)\right) ^{\frac{\nu }{q}}=0,%
\text{ \quad {\rm if\quad }}p.q\text{{\rm -even}}  \label{eq1.35}
\end{equation}
where $l=(N+2)/2$, for $N$-even.

Here we have introduced a generalization of the numbers $d(n)$ of eq.(\ref
{eq1.27}): 
\begin{eqnarray}
d_{k}(n) &=&\frac{\sin (\frac{n(N+1-k)\pi }{N+2})}{\sin (\frac{n\pi }{N+2})}%
,\quad k=0,1,2,...,N-1,  \nonumber \\
\quad n &=&1,2,...,\leq \frac{N+2}{2}  \label{eq1.36}
\end{eqnarray}
which satisfy the following sum rules 
\begin{equation}
d_{i}(n)d_{j}(n)=\sum_{k}(N_{i})_{j}^{k}\ d_{k}(n).  \label{eq1.37}
\end{equation}
From these numbers we observe that the characteristic polynomials of the
fusion matrices of the fields comparing the same cominimal equivalence
classes have equivalent spectra of zeros, i.e., they differ only in the $%
Z_{N}$-degeneracy of their eigenvalues which depend on of the parafermionic
charge through the relation $N=p\nu /q$.

Therefore there are many alternative ways of constructing the $SU(N)_{2}$
polynomial rings in two-variables: Take for $y$ any field belonging to any
equivalence class, $[\phi _{k}^{k}]$. The fusion rules (\ref{eq1.18}) gives
us four possibilities (at most) to choose the field associated with the
variable $x$. The corresponding constraint equations are given by (\ref
{eq1.32}-\ref{eq1.35}). If at least one of the fusion matrices associated
with $x$ and $y$ is non degenerate, it is possible to eliminate one of
variables resulting in a polynomial ring in a single variable.

These results tell us that $SU(N)$, for $N$-old possess a single variable
polynomial ring at level $K=2$. For other values of $K$, as observed by
Gannon \cite{Gannon}, $SU(2)$ and $SU(3)$ are the only $SU(N)$ whose fusion
ring at all level $K$ can be represented by polynomials in only one
variable. For each $N>3$, there will be infinitely many $K$ for which the
fusion ring $SU(N)_{K}$ requires more than one variable, and infinitely many
other $K$ for which one variable will suffice.

\ack{I would like to thank Profs. Roland K\"{o}berle and Angela Forester for
useful discussions.}

\end{document}